\documentclass[letter]{aa} 
\usepackage{natbib}
\usepackage{graphicx}
\usepackage{txfonts}
\usepackage{epsfig}
\usepackage{times}
\usepackage{rotating}
\usepackage{float}
\usepackage[caption = false]{subfig}
\usepackage{setspace}
\usepackage{lscape}
\usepackage{epstopdf}
\usepackage{tabularx}
\usepackage{caption}
\usepackage{xcolor} 

\usepackage{natbib,twoopt}
\usepackage[breaklinks=true]{hyperref} 
\bibpunct{(}{)}{;}{a}{}{,} 
\makeatletter
\newcommandtwoopt{\citeads}[3][][]{\href{http://adsabs.harvard.edu/abs/#3}%
{\def\hyper@linkstart##1##2{}%
\let\hyper@linkend\@empty\citealp[#1][#2]{#3}}}
\newcommandtwoopt{\citepads}[3][][]{\href{http://adsabs.harvard.edu/abs/#3}
{\def\hyper@linkstart##1##2{}%
\let\hyper@linkend\@empty\citep[#1][#2]{#3}}}
\newcommandtwoopt{\citetads}[3][][]{\href{http://adsabs.harvard.edu/abs/#3}%
{\def\hyper@linkstart##1##2{}%
\let\hyper@linkend\@empty\citet[#1][#2]{#3}}}
\newcommandtwoopt{\citeyearads}[3][][]%
{\href{http://adsabs.harvard.edu/abs/#3}
{\def\hyper@linkstart##1##2{}%
\let\hyper@linkend\@empty\citeyear[#1][#2]{#3}}}
\makeatother

\DeclareCaptionFormat{cont}{#1 (cont.)#2#3\par}

\begin{document}


\title{EMIR confirmation of the high-$z$ Ly$\alpha$ blob SHARDS20018464}

\subtitle{}

\authorrunning{S. Zarattini et al.}

\titlerunning{EMIR confirmation of the high-$z$ Ly$\alpha$ blob SHARDS20018464}

\author{S. Zarattini\inst{1}
    \and F. Garzón \inst{2,3}
	\and C. Muñoz Tuñón\inst{2,3}
}

\institute{
    Centro de Estudios de F\'isica del Cosmos de Arag\'on, Plaza San Juan 1, 44001, Teruel, Spain \email{szarattini@cefca.es}
	\and Instituto de Astrofísica de Canarias, IAC, V\'{\i}a Láctea s/n, 38205, La Laguna (S.C. Tenerife), Spain
    \and Departamento de Astrofísica, Universidad de La Laguna, 38206, La Laguna (S.C. Tenerife), Spain
}

\date{\today}
\abstract{}
{A spectroscopic follow-up of the Ly$\alpha$ blob SHARDS20018464 was performed using the infrared spectrograph EMIR (Espectrógrafo Multiobjeto Infra-Rojo) at the 10.4 m GTC telescope, equipped with its newly upgraded H2RG focal plane array detector.}
{We observed the blob in the Ks band, which corresponds, at the photometric redshift of $z=3.49$ presented in the literature, to the rest-frame visible part of the spectrum.}
{Thanks to the sensitivity of EMIR, we were able to identify the [OIII] emission doublet at the position of the blob and the companion galaxy. The flux of the detected lines is of the order of $10^{-17}$ erg s$^{-1}$ cm$^{-2}$, obtained with 4-hour exposure. The mean redshift derived by analysing the two lines individually is $z=3.59\pm0.01$, slightly higher than the photometric one, although in agreement within the uncertainties. We also estimated the flux of the two lines and their ratio, which was found to be $R = 2.92$, very close to the theoretical prediction of 2.98. This is a strong analytical confirmation that the observed emissions correspond to the [OIII] doublet.}
{Within uncertainties, the redshift of the blob confirms the photometric one reported in the literature, as well as the physical properties inferred for it. Its remarkably small size (5–8 kpc vs. $\sim 100$ kpc for typical samples) and its position aside the companion galaxy also confirm the previously-proposed idea that this object could be the first of a new class of compact Ly$\alpha$ emitters. Moreover, the data support the scenario of photo-ionisation due to a cluster of young and massive stars.}

\keywords{Lyman Alpha Blobs, High Redshift Galaxies, }

\maketitle

\section{Introduction}
\label{sec:intro}

High-redshift Ly$\alpha$ emission spans a diverse family of phenomena, whose nomenclature has evolved as observational techniques have improved. At the compact end, Ly$\alpha$ emitters (LAEs) are individual star-forming galaxies selected via their strong Ly$\alpha$ line, historically identified through narrow- or medium-band imaging tomography \citep[e.g.][]{Matsuda2004} and, more recently, through contiguous medium-band photometric surveys such as SHARDS \citep{PerezGonzalez2013,Barro2019,Arrabal2018} and J-PAS/miniJPAS \citep{Rahna2022,Torralba-Torregrosa2023}. Deep integral-field spectroscopy with MUSE has since revealed that most, if not all, star-forming galaxies at $z \gtrsim 3$ are surrounded by extended, low-surface-brightness Ly$\alpha$ halos (LAHs): centrally peaked, exponentially declining envelopes with scale lengths of a few kpc, interpreted as resonant scattering of Ly$\alpha$ photons produced in the central HII regions into the circumgalactic medium \citep{Wisotzki2016,Leclercq2017}. This is distinct from the classical Ly$\alpha$ blobs (LABs) — rare, highly luminous nebulae extending tens to over a hundred kpc, often irregular and multi-component, preferentially found in overdense/protocluster environments and powered by a combination of AGN, starburst-driven outflows, satellite galaxies, or cold accretion \citep{Steidel2000, Nilsson2006,Bacon2021}.

Various mechanisms were proposed as the engine of Ly$\alpha$ blobs. In particular, the gas can be excited by obscured AGN \citep[e.g.][]{Miley2008}, starburst galaxies \citep[e.g.][]{Wilman2005} or cooling radiation from cold gas streams \citep[e.g.][]{Nilsson2006}, although the latter mechanism is disfavoured in other publications. \citet{Prescott2015} and \citet{Ao2020} suggested that pure cooling radiation is not able to explain other properties observed in their object, such as line ratios and kinematics, thus disfavouring this scenario as the dominant mechanism for their objects.

The giant LABs are mostly found in high-density regions. \citet{Steidel2000} suggested that they are tracers of the formation of the most-massive galaxies in the cores of massive protoclusters and \citet{Matsuda2004} extended their results, finding 33 more LABs in the same field. About 90\% of their LABs are found in the same over densities defined in \citet{Steidel2000}.

Recently, \citet{Zarattini2025} presented the discovery and characterisation of a LAB detected within the SHARDS survey. The photometric redshift of the blob and companion galaxy was estimated in $z = 3.49 \pm 0.06$ \citep{Barro2019,Arrabal2020}. Moreover,
\citet{Zarattini2025} showed that the photo-ionisation source for this blob is compatible with a cluster of massive stars located in the companion galaxy, that is able to ionise the gas of the blob. According to these authors, the LAB is particularly small, with a size of $5-8$ kpc, smaller than the usual LABs, which are typically $\sim 100$ kpc in size. This suggests a possible new class of compact LABs -- small, luminous, and not associated to a galaxy -- making this target of particular interest and possibly disconnected from the protocluster scenario. However, a complete study of the environment of this LAB is still pending.

In this work we use the Espectr\'ografo Multiobjeto Infra-Rojo (EMIR) of the GTC to obtain a spectroscopic confirmation of the redshift of the blob and companion galaxy. In Sect. \ref{sec:data} we describe the observations and data reduction. In Sect. \ref{sec:results} we present the results of our analysis, in particular the detection of the [OIII] emission lines and we compute the spectroscopic redshift and the fluxes of the two components of the doublet. Finally, in Sect. \ref{sec:discussion} we discuss our results and draw our conclusions.

The cosmology used in this paper is H$_0 = 69.6$, $\Omega_m = 0.286$ and $\Omega_m = 0.714$ \citep{Bennett2014}.

\section{Data}
\label{sec:data}
The LAB, SHARDS20018464, was observed in March 2026 with the EMIR instrument \citepads{2022A&A...667A.107G} on the 10.4-m GTC telescope. EMIR was recently upgraded with a new H2RG focal plane array (FPA) detector, resulting in a significant increase in sensitivity and a substantial reduction of the detector instabilities that had previously limited the instrument's performance.

\begin{figure}
        \begin{center}
        \includegraphics[width=\columnwidth]{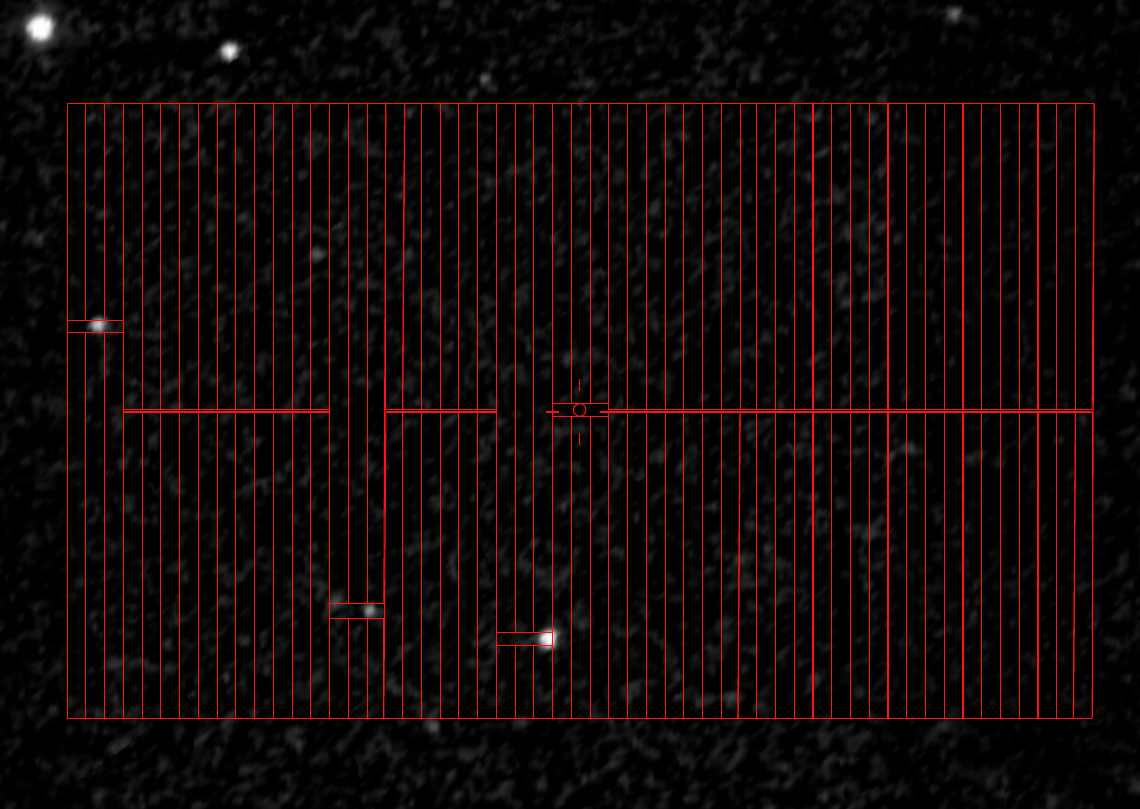}
        \caption{EMIR MOS mask used for the observations. The multislit in the centre was in the position of the blob. Three reference stars were also included. From top to bottom: GOODS J123709.52+621233.5, $Kmag=15.395$; TKRS10556, $Kmag=16.962$; GOODS J123734.79+621435.0, $Kmag=16.113$. Magnitudes are in AB system. North is up and East, left. The EMIR mask dimensions are $4\arcmin\times6.67\arcmin$.}
        \label{fig:mask}
        \end{center}
\end{figure}

Six standard ABBA nodding cycles with 600s integration time on each pointing nod were taken with EMIR configured in spectroscopic mode in the K band, totalling 4 hours of observation. A MOS mask was specifically designed for this observation using the EMIR  Optimizer Slit Positioner utility \citepads[OSP,][]{2022A&A...667A.107G} that accommodated the LAB in the central portion plus three additional reference stars used for fine pointing and photometric calibration. Fig. \ref{fig:mask} shows the mask overlayed onto a 2MASS image in the J band.

Once acquired, the data were reduced using the dedicated EMIR pipeline, PyEMIR \footnote{\url{https://github.com/guaix-ucm/pyemir}} \citepads{2019ASPC..523..317C, 2019hsax.conf..605C, 2019ASPC..521..232P}, after individually inspecting all frames to identify any potential artefacts or undesirable effects. Although the new H2RG detector allows for per-frame bias correction using the first and last four detector rows, this correction was not applied, as the data quality and background uniformity were found to be excellent. Analysis of the 32 readout channels demonstrates that temporal drift in reference-row averages is smaller than their statistical uncertainty. Consequently, local per-channel bias subtraction introduces net noise rather than reducing error.

PyEMIR corrects for a range of instrumental effects and ultimately produces sky-subtracted, wavelength-calibrated two-dimensional spectra. Wavelength calibration is performed using the numerous night-sky emission lines present in the data. Prior to this step, the spectra are rectified to remove the intrinsic optical geometric curvature using a dedicated set of calibration frames acquired after EMIR was installed at the GTC.

\begin{figure}
        \begin{center}
        \includegraphics[width=\columnwidth]{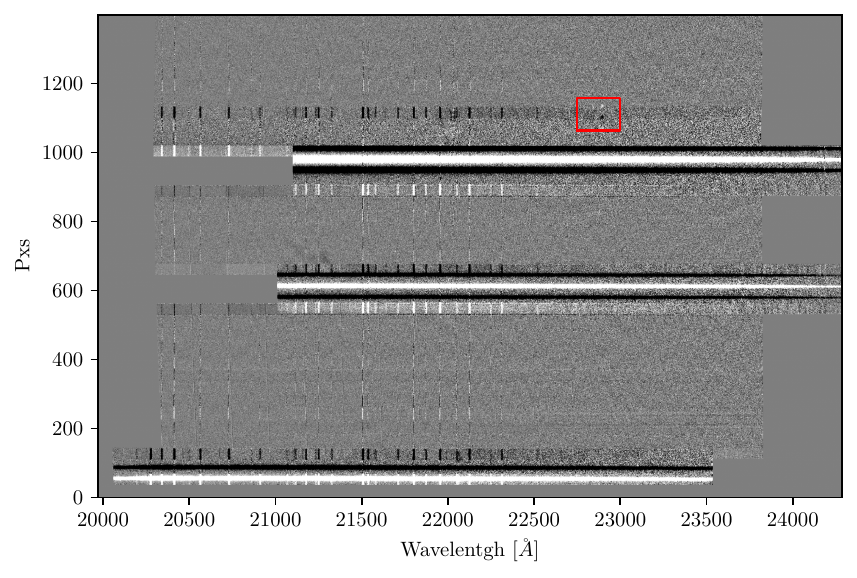}
        \caption{Final reduced image, showing the spectra of the three reference stars and, in red rectangle, the location of the [OIII] lines of SHARDS20018464. Note that spectral coverage differs for each of the stars, due to their different position in the EMIR mask.}
        \label{fig:final}
        \end{center}
\end{figure}

Due to the long exposure times of individual pointing, the intensity of the sky emission lines varies between frames. As a result, residual sky features remain after image combination and subtraction, and can be seen in the final reduced data. However, no sky emission lines fall within the wavelength range of interest for this study. The final reduced image is shown in fig. \ref{fig:final}. The spectra of the three reference stars are clearly visible, whereas no continuum emission is detected in the region of the LAB. The red rectangle indicates the location of the two [OIII] emission lines of SHARDS20018464. This region is shown in greater detail in fig. \ref{fig:zoom}, where the positive and negative signals, arising from the ABBA nodding pattern, are highlighted with red and green circles, respectively.

The line connecting the positions of the two peaks, corresponding to the two [OIII] emission lines, is slightly tilted with respect to the detector rows, which in principle define the direction of spectral dispersion. This effect may be due to deviations in the mask alignment on the sky, residuals in the spectral rectification process, or, more likely, jitter of the GTC during each of the long individual exposures. This introduces an additional source of uncertainty in the determination of the spectral position of each line.

After line identification, the spectrum of interest was extracted by summing 12 rows around the peak position. This width is set by the median seeing during the night ($1\arcsec$) and by measuring the spatial extent of the emission in each of the three reference stars. 
{Based on the spatial FWHM of three reference stars ($1.0\arcsec-1.1\arcsec$ seeing; $\sim$5 pixels), the optimal extraction aperture to maximize SNR was empirically set to 10 pixels. We adopted a slightly wider 12-pixel aperture to capture additional flux, confirming that high-accuracy local background subtraction leaves the net flux unaffected.

The resulting spectrum was then convolved with a 6-pixel boxcar, corresponding to the slit width ($1.2\arcsec$). Finally, the spectral pseudo-continuum was modelled using a bicubic spline fitted to a 100-pixel-wide running median covering the full spectral range, and subsequently subtracted from the original spectrum.

Figure \ref{fig:spec_lines} shows the spectral region containing the two [OIII] lines. The red vertical lines mark the positions of the peaks of the Gaussian fits to each line, which are adopted as the measured wavelengths from which the redshift is derived in each case. Several factors contribute to the uncertainty in the wavelength measurement. The first is the intrinsic spectral calibration using sky lines, which PyEMIR estimates to be 0.482 \AA \,in this case. The second is the uncertainty in the line centroid determination, which scales as FWHM/SNR, with FWHM values of 10.811 \AA\ and 14.558 \AA\ for each respective line ([O III] $\lambda 4959$ and $\lambda 5007$) and SNRs, measured from the extracted spectrum, are 8.7 and 5.3, respectively. The latter is combined in quadrature with the statistical uncertainty from the Gaussian fit (0.358 \AA\ and 0.186 \AA). Finally, spectral dispersion induced by guiding variations during the integration contributes an additional error term, which dominates in this case.  The guiding error has an RMS of 0.65", which corresponds to 11.496 \AA\ given EMIR’s pixel scale of 0.1945"/pix and a spectral dispersion of 3.44 \AA/pix. Since the final spectrum results from the combination of 24 individual exposures (6 ABBA cycles), this error scales with the square root of 24, yielding a value of 56.32 \AA, which completely dominates the final error budget in quadrature. Adding these terms in quadrature yields a final uncertainty in the redshift of 0.01 for each line.

\begin{figure}
        \begin{center}
        \includegraphics[width=\columnwidth]{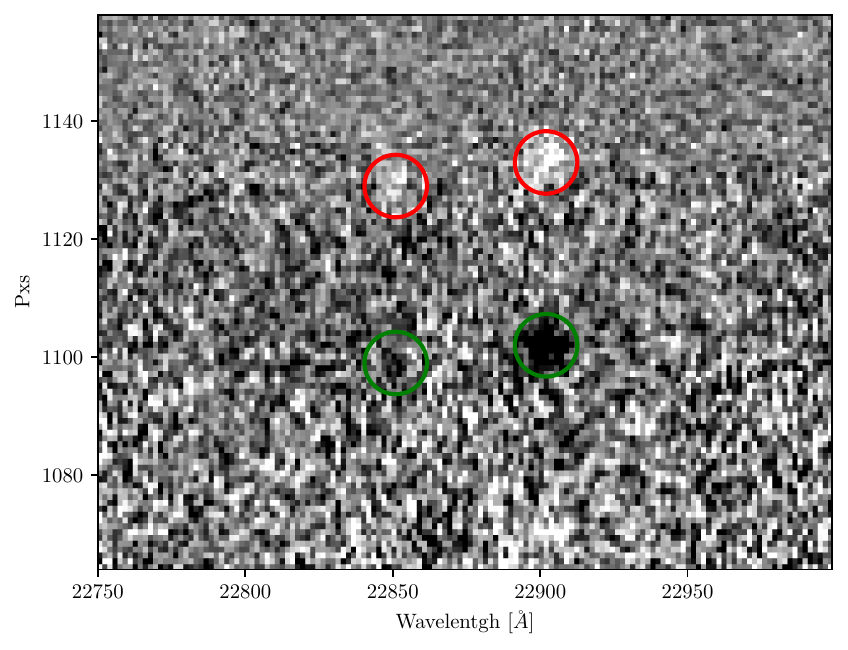}
        \caption{Zoomed view of the detector area where the [OIII] lines are located. Red circles mark the position of each detection and are used for subsequent measurement. Green circles are the residuals of the data reduction, with negative count values.}
        \label{fig:zoom}
        \end{center}
\end{figure}

For the flux measurement, the sensitivity of the final image was calibrated using the total counts of the reference stars. As noted above, the spectral coverage of each star is not identical; however, no explicit correction factor was applied, and the average of the three individual calibration factors was adopted. It is difficult to estimate the uncertainty in the photometric calibration, which in any case affects both lines equally, and is unlikely to be smaller than 20\%. Note that while the reported flux uncertainties in table \ref{tab:results} account for the statistical noise propagation in the extraction process, the final photometric budget is dominated by the systematic variance associated with the flux conversion factor. The 20\% value is estimated from the ratio of the standard deviation to the mean of these conversion factors translated to flux. Taking the maximum peak-to-peak difference between individual factors rather than the standard deviation yields a ratio of 48\%, which serves as a conservative upper limit for the error in fluxes.

\section{Results}
\label{sec:results}
In Fig. \ref{fig:spec_lines} we show the spectrum at the position of our target, centred on the position of the two [OIII] lines. Due to the very faint magnitude of the source, that was estimated to be $\sim 27$ AB mag from Hubble Space Telescope (HST) observations in the F606W filter \citep{Zarattini2025}, the continuum is not visible in our data. We found a slightly different redshift for each of the lines, namely $z=3.61$ for the bluest one ([OIII] at 4959 \AA) and $z=3.57$ for the reddest one ([OIII] at 5007 \AA).

We were also able to measure the integrated fluxes of the two detected lines, that are $9.21 \times 10^{-18}$ erg s$^{-1}$ cm$^{-2}$ and $2.69 \times 10^{-17}$ erg s$^{-1}$ cm$^{-2}$ for the 4959 and 5007 \AA\, lines, respectively. The two lines are clearly detected, with a signal-to-noise ratio SNR $\sim 5.3$ and SNR $\sim 8.7$ for the bluest and reddest line, respectively. The ratio of these two lines is 2.92, very close to the theoretical ratio of 2.98 that is expected for the [OIII] doublet \citep{Storey2000}. This intensity ratio is a strong confirmation that the observed lines are the [OIII] doublet lines of our target. The results of our analysis are reported in Table \ref{tab:results}.

\begin{figure}[t]
        \begin{center}
        \includegraphics[width=\columnwidth]{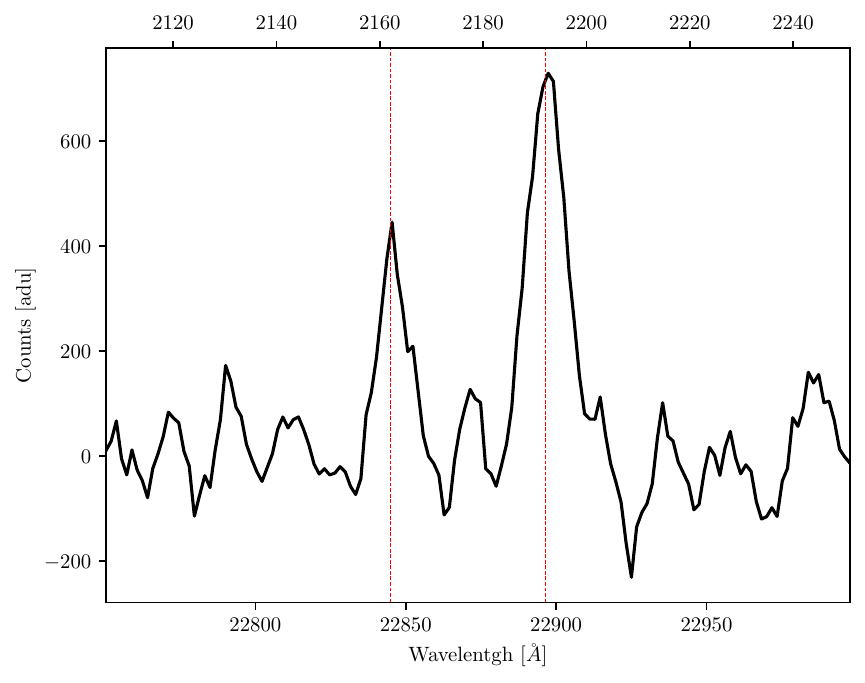}
        \caption{Portion of the final spectrum  where the two [OIII] lines are located. Red vertical lines are in the peaks of the Gaussian fit to each line.}
        \label{fig:spec_lines}
        \end{center}
\end{figure}

\section{Discussion and conclusions}
\label{sec:discussion}
The goal of this work was the spectroscopic confirmation of the redshift of the LAB called SHARDS20018464, which photometric redshift was $z_{phot} = 3.49\pm0.06$ \citep{Barro2019,Arrabal2020}. Due to the extremely faint magnitude of the object \citep[27 mag in the HST F606W filter,][]{Zarattini2025}, observations proved highly complicated. Indeed, a tentative detection of emission lines within the blob's spectrum was achieved only by using a 10-meter-class telescope equipped with an IR instrument featuring the sensitivity of EMIR, recently equipped with its newly upgraded H2RG focal plane array detector. The detection of such faint lines, with fluxes as low as $10^{-17}$ erg s$^{-1}$ cm$^{-2}$, in just 4 hours of integration would not have been possible with EMIR equipped with its original detector.

Our results strongly support the high redhift nature of SHARDS20018464. Its photometric redshift was found to be $z=3.49\pm0.06$, whereas our mean spectroscopic redshift is $z=3.59\pm0.01$, placing the LAB slightly farther away. The scale reported in \citet{Zarattini2025} was  7.478 kpc/\arcsec. With the new redshift value the scale become  7.402 kpc/\arcsec, resulting in a separation of 5.6 kpc between the blob and the companion galaxies \citep[it was 5.7 kpc in][]{Zarattini2025}), with the blob size that does not change, since in \citet{Zarattini2025} only a range of $5-8$ kpc was given, that remains valid also with the new redshift. This LAB thus remain smaller than the majority of the known objects of these kind and it could represent the first of a new class of objects.
It is worth noting that \citet{Arrabal2018} found a sample of small, compact, and luminous Ly$\alpha$ blobs. Only for SHARDS20018464 we were able to identify the ioniziong source -- a galaxy at a distance of 5.6 kpc. For the rest of the sample, HST data were not deep enough. 
In future work, we plan to explore the JWST JADES DR5 dataset to identify additional ionizing counterparts and provide further support for the new class proposed for SHARDS20018464.

\begin{table}[t]
    \centering
        \caption{Spectroscopic redshifts and fluxes for the detected emission lines.}
    \begin{tabular}{ccc}
    \hline\hline
    \multicolumn{3}{c}{\rule{0pt}{4ex}} \\ [0.5ex]
    line & $z$ & Flux\\
    \AA &  & erg s$^{-1}$ cm$^{-2}$ \\
    \hline
    \rule{0pt}{3ex} [OIII] 4959 & $3.61\pm0.01$ & $(9.21\pm0.929) \times 10^{-18}$ \\ [0.5ex]
    \rule{0pt}{3ex} [OIII] 5007 & $3.57\pm0.01$ & $(2.69\pm0.096) \times 10^{-17}$ \\ [0.5ex]
    \hline
    \end{tabular}
    \label{tab:results}
\end{table}

\begin{acknowledgements}
This work is part of the collaboration ESTALLIDOS, supported by the Spanish research grants, PID2022-136598NB-C31 and by the Government of the Canary Islands through EU FEDER funding project PID2021010077. This work has also been supported by the Agencia Estatal de Investigación del Ministerio de Ciencia e Innovación (MCIN/AEI) under grant PID2021-129031NB-I00 and the European Regional Development Fund (ERDF) with reference 10.13039/501100011033. S.Z. acknowledges the financial support from the Spanish Ministry of Science and Innovation and the European Union - NextGenerationEU through the Recovery and Resilience Facility project ICTS-MRR-2021-03-CEFCA; the Governments of Spain and Arag\'on through their general budgets; the Fondo de Inversiones de Teruel, and the Aragonese Government through the Research Group E16\_23R. Based on observations made with the Gran Telescopio Canarias (GTC), installed at the Spanish Observatorio del Roque de los Muchachos of the Instituto de Astrofísica de Canarias, on the island of La Palma. This work is based on data obtained with the instrument EMIR, built by a Consortium led by the Instituto de Astrofísica de Canarias. EMIR was funded by GRANTECAN and the National Plan of Astronomy and Astrophysics of the Spanish Government. This work is part of grant CEX2025-001609-S, awarded to the Instituto de Astrofísica de Canarias under the Severo Ochoa Centre of Excellence program and funded by MICIU/AEI/10.13039/501100011033.

\end{acknowledgements}

\bibliography{bibliografia}

\end{document}